\documentclass[structabstract]{aa}  
\usepackage[]{natbib}
\usepackage{lscape}
\usepackage{longtable}
\usepackage{float}
\restylefloat{figure}

\usepackage{float}
\usepackage{cuted}

\usepackage{graphicx}
\usepackage{amsmath}
\usepackage{txfonts}

\begin{document}
   \title{Extraplanar H{\sc II} Regions in the edge-on Spiral Galaxies NGC~3628 and NGC~4522\thanks{Based on observations gathered as part of observing program 64.N-0208(A), 3.6m telescope with EFOSC2 at ESO, La Silla observatory.}}

    \author{Y. Stein,
          \inst{1}
          D. J. Bomans,
          \inst{1,2}
          A. M. N. Ferguson
          \inst{3}
          \and
          R.-J. Dettmar
          \inst{1,2}
          }

   \institute{Astronomisches Institut (AIRUB), Ruhr-Universit\"at Bochum,
              Universit\"atsstrasse 150, 44801 Bochum, Germany\\
              \email{stein@astro.rub.de}
              \and
              Research Department: Plasmas with Complex Interactions, Ruhr-Universit\"at Bochum,
              Universit\"atsstrasse 150, 44801 Bochum, Germany
              \and
              Institute for Astronomy, University of Edinburgh, Blackford Hill,	Edinburgh EH9 3HJ, UK}

   \date{Received February 09, 2017; Accepted May 30, 2017}

  \abstract 
  {Gas infall and outflow are critical for determining the
  star formation rate and chemical evolution of galaxies but direct
    measurements of gas flows are difficult to make.  Young massive stars and
  H{\sc ii} regions in the halos of galaxies are potential tracers for
  accretion and/or outflows of gas.}  
  {Gas phase abundances of three H{\sc ii} regions in the lower halos of the
  edge-on galaxies NGC~3628 and NGC~4522 are determined by analysing
  optical long-slit spectra. The observed regions have projected
  distances to the midplane of their host from 1.4 to 3 kpc.}  
  {With the measured flux densities of the
  optical nebular emission lines, we derive the oxygen abundance
  12~+~log(O/H) for the three extraplanar H{\sc ii} regions. The
  analysis is based on one theoretical and two empirical strong-line
  calibration methods.}  
   {The resulting
  oxygen abundances of the extraplanar H{\sc ii} regions are
  comparable to the disk H{\sc ii} regions in one case and a little
  lower in the other case. Since our results depend on the accuracy of
  the metallicity determinations, we critically discuss the difference
  of the calibration methods we applied and confirm previously noted
  offsets.  From our measurements, we argue that these three 
  extraplanar H{\sc ii} regions were formed in the disk or at least
  from disk material. We discuss the processes that could transport disk
  material into the lower halo of these systems and conclude that 
  gravitational interaction with a companion galaxy is most
  likely for NGC~3628 while ram pressure is favoured in the case of NGC~4522.}
  {}

\titlerunning{Extraplanar H{\sc ii} Regions in edge-on Spiral Galaxies}
\authorrunning{Stein et al.}

\maketitle

\section{Introduction}
Star formation in spiral galaxies usually takes place in  dense regions of the interstellar medium (ISM), close to the midplane of the disk. However, there is evidence for low-level star formation high above the  midplane in the Milky Way as well as in some external galaxies. OB stars exist at large distances from the plane in the Milky Way \citep[e.g.,~][]{GreensteinSargent74, Kilkennyetal95} as well as in the halos of other galaxies \citep[e.g.,~][]{comeron2001}. Furthermore, extraplanar H{\sc ii} regions were detected in galaxies with strong indication for an exchange of matter between the disk and the halo, e.g., NGC\,55 \citep{ferguson1996}, NGC\,891 and NGC\,4013 \citep{howksavage1999}, as well as the Virgo galaxy NGC~4402 \citep{corteseetal2004}. In some objects, such extraplanar H{\sc ii} regions are found at even larger galactocentric distances and hence are attributed to the intracluster or 
intergalactic medium of interacting galaxies, where the H{\sc ii} regions probably formed in tidal debris \citep{mendesetal2004, ryanweberetal2004}, or by ram pressure stripping, e.g. in the Virgo cluster \citep{gerhardetal2002,oosterloovangorkom2005}. \\ 
\indent There are different scenarios to explain the presence of young stars and star formation in these unusual environments. (I) For the young stars, ejection from the disk by either asymmetric supernovae of a close binary system \citep[e.g.,~][]{stone1982} or dynamical three-/ four-body encounters in dense stellar systems \citep[e.g.,~][]{poveda1967} could be responsible for their existence far from the galactic midplane. Supersonic space motions of massive stars are possible with this mechanism, as apparent from bow shocks induced by these stars \citep[e.g.,~][]{gvbomans2008}.
(II) Star formation in the halo itself would not only explain the
presence of young stars in the halos \citep[e.g.,~][]{keenanetal1986}
but also explain the existence of H{\sc ii} regions in the halo,
formed with halo material. (III) Tidal forces could induce clumps in
material drawn out of galactic disks or in gas falling back onto a
disk \citep[e.g.,~][]{barneshernquist96}, which then would form stars
 and thus explain extraplanar H{\sc ii}
regions.\\ 
\indent One way to constrain the origin of the above
mentioned star formation in galactic halos is the determination of
metallicities of the extraplanar and disk H{\sc ii} regions. Low
metallicities of extraplanar regions in comparison to high metallicity
disk regions would indicate that the region is formed by low
metallicity halo gas (second scenario). Comparable metallicities
between disk and extraplanar regions would lead to the conclusion that
the extraplanar H{\sc ii} region is formed by disk material (first and
third scenario). \\ 
\indent \citet{tuellmann2003} studied extraplanar H{\sc ii} regions in the galaxy NGC~55. They found a lower metallicity of the extraplanar regions in comparison to H{\sc ii} regions in the midplane and concluded that the second scenario is most likely responsible, with the H{\sc ii} regions having formed from halo material. This result was verified by \citet{kudritzkietal2016}. The metallicities of the extragalactic H{\sc ii} regions in Stephan's Quintet \citep{mendesetal2004}, in the interacting galaxies studied by \citet{ryanweberetal2004} and in the extraplanar H{\sc ii} regions in the Virgo galaxy NGC~4402 are all enhanced in metallicity. In the case of the Virgo galaxy NGC~4388, subsolar metallicities are observed. This suggests there may be no single mechanism for producing extraplanar H{\sc ii} regions. However, there have only been a limited number of studies so far and the nature and origin of such  objects requires further study. \\ 
\indent NGC 3628 in the Leo Triplet and NGC 4522 in the Virgo Cluster are two edge-on galaxies. They are known to have extended diffuse H$\alpha$ structure, which we refer to as extraplanar Diffuse Ionized Gas (eDIG) \citep{rossadettmar2000A&A}. Three extraplanar H{\sc ii} regions are known in these systems. We here present optical long-slit spectroscopy of these regions which we analyse to determine metallicities and thus place further constraints on the origins of the young stars and gas at large scaleheights.\\ 
\indent The extraplanar H{\sc ii} region (1) of NGC 4522 is offset from the main star-forming region defined by H$\alpha$ and lies at a projected distance of 1.4 kpc. This is also visible in the HI map of NGC~4522 \citep{kenney2004}. According to \citet[][Fig. 4]{kenney2004} region (1) resides in extraplanar gas that was stripped out and removed from the disk via ram pressure. The extraplanar H{\sc ii} regions (2) and (3)  lie in NGC~3628 and have projected distances from the midplane of their host galaxy of 2.8 kpc and 3.0 kpc, respectively. We use the oxygen abundance 12~+~log(O/H) to derive metallicities. Oxygen is widely used as a reference element because it is relatively abundant, emits strong lines in the optical (as an important cooling element), and is observed in many ionization states \citep{kewleydopita2002}. With the aid of excitation models, oxygen abundances can also be used to derive electron temperatures. There are different ways to derive oxygen abundances and temperatures. Direct methods use ratios of auroral and nebular lines as they are temperature sensitive (e.g.,~ the auroral line [OIII] $\lambda$4363 in combination with [OIII] $\lambda$$\lambda$4959, 5007). The line intensities of the auroral lines in spectra are weak, especially at solar metallicities or higher. We did not detect any of these weak lines. Our spectra contain the "strong emission lines" like the Balmer lines, [OII] $\lambda$$\lambda$3726, 3729, [OIII] $\lambda$$\lambda$4959, 5007, [NII] $\lambda$$\lambda$6548, 6583. Due to the spectral resolution, the lines of [O II] $\lambda$3726.0 and [O II] $\lambda$3728.8 were blended in the spectra and so we use the sum of the doublet lines. To derive oxygen abundances, we use strong-line calibrations. There are different strong-line calibration types, empirical \cite[e.g.,~][]{pilyugin2001, pilyuginthuan2005, pettinipagel2004, pilyuginetal2014} and theoretical \cite[e.g.,~][]{kewleydopita2002, KK2004}, which show well-known offsets between 0.1 and 0.7 dex \citep{stasinska2002, modjazkewley2008, moustakas2010}. The empirical methods use a fitted relationship between direct metallicities and strong-line ratios in the optical of a sample of H{\sc ii} regions to derive metallicity relations while the theoretical methods use photoionization models like Cloudy \citep{ferland1998} to predict how theoretical emission line ratios depend on the input metallicity. We will use three different calibrations, investigate their offsets, and discuss possible reasons. Using our
measurements of the oxygen abundances, we explore the origin of the extraplanar H{\sc ii} regions. \\ 
\indent The data reduction and observational strategy is given in Section 2. We then describe the measurement of the emission lines and the different calibration types for the analysis in Section 3. Section 4 contains the results. In Section 5 a short summary and the conclusions are given.    

\section{Data}

\subsection{Spectroscopic data}
\begin{figure*}
\begin{center}
\sidecaption
{\resizebox{13cm}{!}{\includegraphics{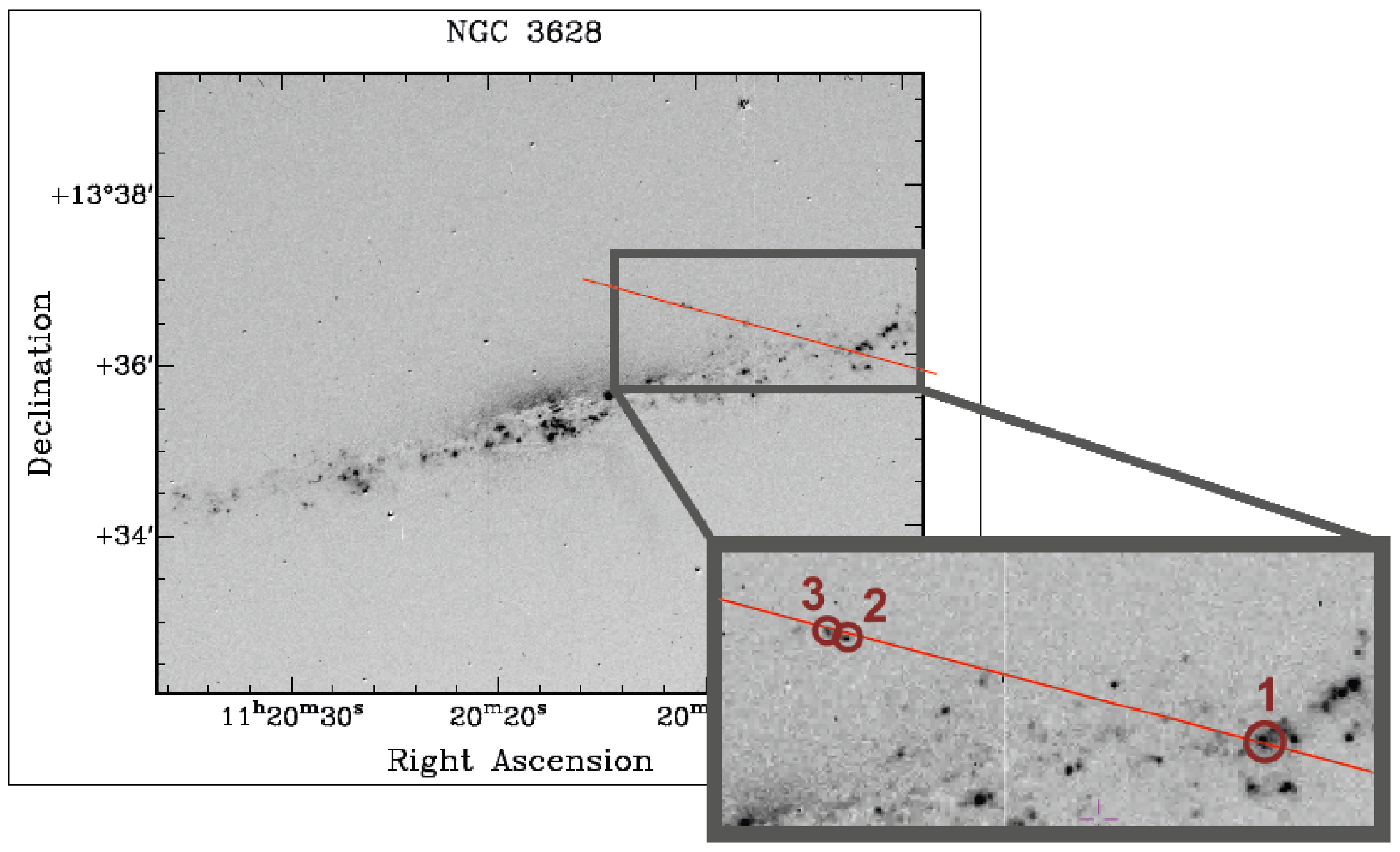}}}
\caption{Slit position on an H$\alpha$ image of NGC~3628. The marked regions are analyzed.}
\label{NGC 3628}
\end{center}
\end{figure*}

\begin{figure*}
\begin{center}
\sidecaption
{\resizebox{13cm}{!}{\includegraphics{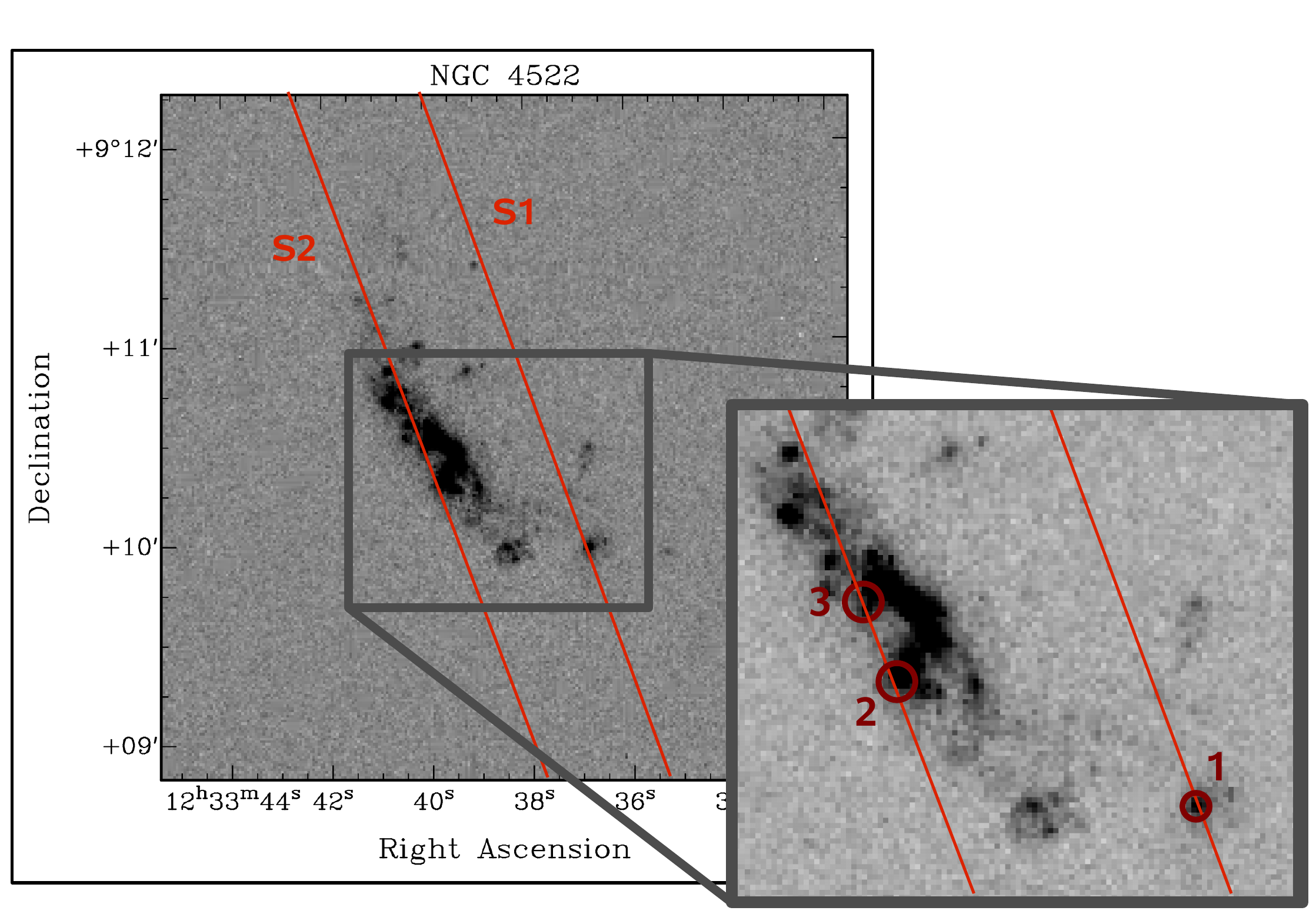}}}
\caption{Slit position on an H$\alpha$ image of NGC~4522. The marked regions are analyzed.}
\label{NGC 4522}
\end{center}
\end{figure*}

\begin{figure*}
	\centering
		\includegraphics[width=1.0\textwidth]{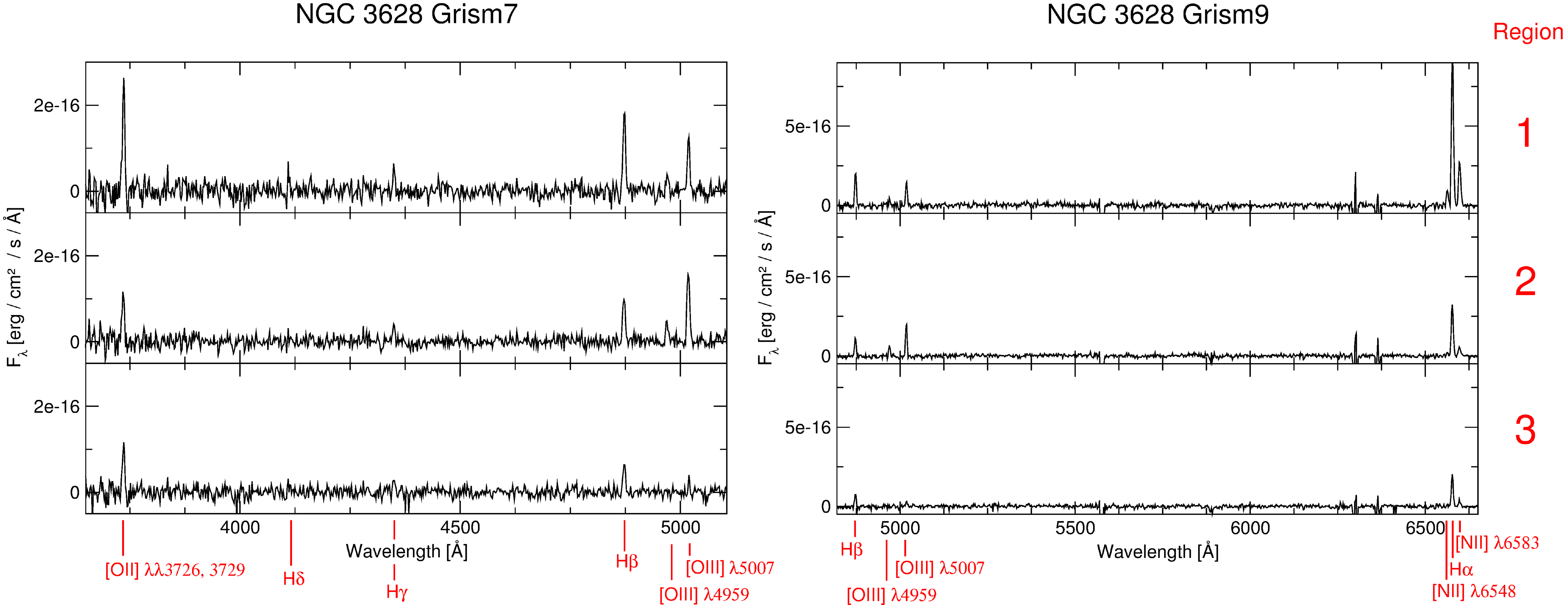}
\caption{Spectrum of NGC~3628 (grism 7 and grism 9).}
\label{Spektrum 9 NGC 3628}
\label{Spektrum 7 NGC 3628}
\end{figure*}

\begin{figure*}
	\centering
		\includegraphics[width=1.0\textwidth]{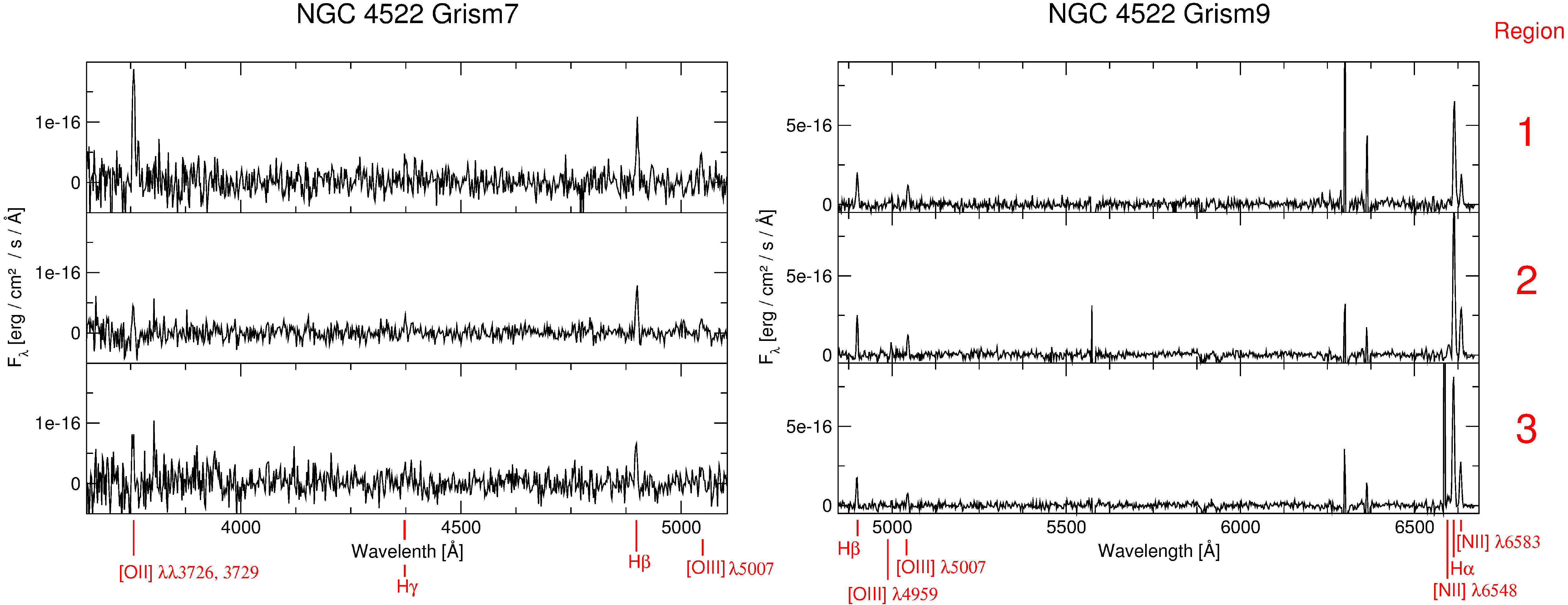}
\caption{Spectrum of NGC~4522 (grism 7 and grism 9).}
\label{Spektrum 9 NGC 4522}
\label{Spektrum 7 NGC 4522}
\end{figure*}

The spectroscopic data were obtained at the ESO La Silla observatory with EFOSC2 (ESO Faint Object Spectrograph and Camera (v.2)) attached to the 3.6 m telescope during the nights of March 7 -- 9, 2000. A 2048 $\times$ 2048 CCD detector with a pixel size of 15 $\mu$m was used. The field of view is 5.4\arcmin x 5.4\arcmin with a pixel scale of 0.316\arcsec/pixel. With the two different grisms 7 (3270 -- 5240 \AA) and 9 (4700 -- 6700 \AA), we were able to detect the important lines in the optical. The measured spectral resolution is 6.7 \AA \ in both grisms. The slit had a length of 5\arcmin \ and a width of 1\arcsec. The seeing was around 1.1\arcsec (see Table \ref{tab:4Galaxien}). The long slit was oriented differently in the galaxies. In Fig. \ref{NGC 3628} and Fig. \ref{NGC 4522} the slit positions for the galaxies NGC~3628 and NGC~4522 are shown and in Table \ref{tab:4Galaxien} we report the position angles. The slit position was chosen to go through both extraplanar regions and the disk. The total integration time for NGC~3628 was one hour in each grism. For NGC~4522, two slit positions were observed with the first covering the extraplanar region (s1) and the second the galaxy (s2). The total integration time was 20 min in slit position s1 with grism 7 and slit position s2 with both grisms. Slit position s1 with grism 9 was observed with two exposures of 20 min each (see Table \ref{tab:4Galaxien} for details of the observations). \\ 
\indent The data reduction was carried out using IRAF \citep{tody1986, tody1993}. This includes the corrections for bias and overscan, flat-field response and cosmic rays with L.A.Cosmic \citep{vandokkum2001}. Additionally, the wavelength calibration was achieved with Helium-Argon spectra and the flux calibration was done with standard stars (EG274, LTT2415). In order to obtain the pure emission from the H{\sc ii} regions, night sky background emission above and below these regions was averaged and subtracted with the IRAF task {\tt background}. There are some residuals at 6300 \AA \ in the spectrum of NGC~4522 grism 9 left, which do not influence the data analysis. The galaxy continuum was subtracted by the IRAF task {\tt continuum}, which was used with a linear spline 12th order and finally the spectra were coadded. Fig.~\ref{Spektrum 7 NGC 3628} - Fig.~\ref{Spektrum 9 NGC 4522} show the resulting spectra. To correct for stellar absorption, the non-continuum-subtracted spectra were coadded. All listed flux spectral densities are taken from these non-continuum-subtracted spectra.
    
\subsection{Narrow-band imaging}
The H$\alpha$ imaging for the galaxies discussed in this paper was generated from two different sources:
NGC~3628 was observed on May 8, 1991 using the ESO NTT (red arm of the EMMI instrument) 
for 900s in H$\alpha$ and  300s in R as part of ESO program 047.01-003.
The detector was a Ford $2048^2$ single CCD (ESO CCD \#24). Both images were taken with subarcsecond seeing. The data 
and all relevant calibration files taken during the run and a few days before and after the observing run were retrieved 
from the ESO archive and re-reduced by us using IRAF in the usual manner. L.A.Cosmic \citep{vandokkum2001} was used to 
clear the image of cosmic rays to produce the final H$\alpha$ and final R image. 
The continuum subtracted image was then produced by aligning H$\alpha$ and R band, homogenizing the PSF
and subtracting the appropriately scaled R image from the H$\alpha$ image \citep[e.g.,~][]{Skillman1997}.  
To determine the scaling factor, we measured the apparent fluxes for several stars in the field on both the 
H$\alpha$ and R images. Scaling, subtraction and correction to reach an optimal correction of the continuum 
light of the galaxy was performed with our own IRAF scripts.
The NGC~4522 H$\alpha$ data were taken directly from the work of
\citet{Koopmann2001} and the GOLDMINE data base \citep{Gavazzi2003}.
The images were astrometrically calibrated to the DSS system.
Flux calibration of the continuum subtracted H$\alpha$ images was performed by
transferring the total SDSS r' band flux of each galaxy to the its continuum
image. With the derived scaling factor (see above), the flux scale can then be
directly transferred to the continuum corrected H$\alpha$ image.  A conversion
from magnitudes to flux is performed using the fact that SDSS is calibrated in AB
magnitudes, so that the zero-point flux density of each filter is 
3631~Jy (with 1~Jy = 1~Jansky = 10$^{-26}$ W Hz$^{-1}$ m$^{-2}$ = 10$^{-23}$ erg s$^{-1}$ Hz$^{-1}$ cm$^{-2}$)\footnote{http://classic.sdss.org/dr7/algorithms/fluxcal.html}.

\tiny
\begin{table*}
\caption{Galaxy parameter and observation log.}
\begin{tabular}[ht]{cccccccccccc}
\hline\hline 
Galaxy & R.A. (J2000)            & Dec (J2000)           & Hubble type  &     $v$ [km/s]  &  D  [Mpc] & i [$^\circ$] & Grism\#   & t$_{int}$ [s] & PA & Seeing & AM \\
         &    [1]                 &     [2]               &   [3]        &  [4]        & [5]       &     [6]       &  [7]         & [8]         & [9] & [10]& [11] \\ [0.2em]  \hline
NGC~3628 & 11$^h$20$^m$17$^s$& 13$^{\circ}$35'23\arcsec& Sb pec& 843$\pm$1 & 12.2$\pm$2.4 &  87\tablefootmark{(1)}  & 7      &  2x1800  & -15.50° & 1.11  & 1.40 \\
         &										&									&									 			&			&		  				&           &9			&	2x1800 	&		& 1.11 &	1.38\\

NGC~4522 &12$^h$33$^m$39.7$^s$ &09$^{\circ}$10'30\arcsec& SB(s)cd  & 2329$\pm$1 & 16.8$\pm$1.3 & 75 $\pm$ 5\tablefootmark{(2)} & 7 s1  &  1x1200 & -70.92° &  n.a. & 1.34\\
         &										&									&						&									&									&					& 9	s1	&	2x1200  	&		&	 n.a. & 1.34\\
         &										&									&						&									&									&					& 7 s2	&	1x1200  	&	-70.90°	&	 n.a. & 1.88 \\
         &										&									&						&									&									&					& 9	s2	&	1x1200  	&		&		 n.a. & 1.71\\
         \hline
\end{tabular}
\tablefoot{Right Ascension [1] and Declination [2] of the galaxies from NED. [3] Hubble type from \cite{vau91}. Radial velocity [4], and distance [5] (if available averaged distances are used) were taken from NED. [6] is the inclination, [7] the grism number as well as the slit number (s1, s2 for NGC~4522), [8] shows the integration time, [9] is the position angle of the slit, [10] is the seeing (not available for NGC~4522), and [11] is the airmass.\\
\tablefoottext{1}{\citet{tully}}, \tablefoottext{2}{\citet{kenney2004}}
}
\label{tab:4Galaxien}
\end{table*}
 \normalsize

\begin{figure*}
   \centering
   \includegraphics[width=1.01\textwidth]{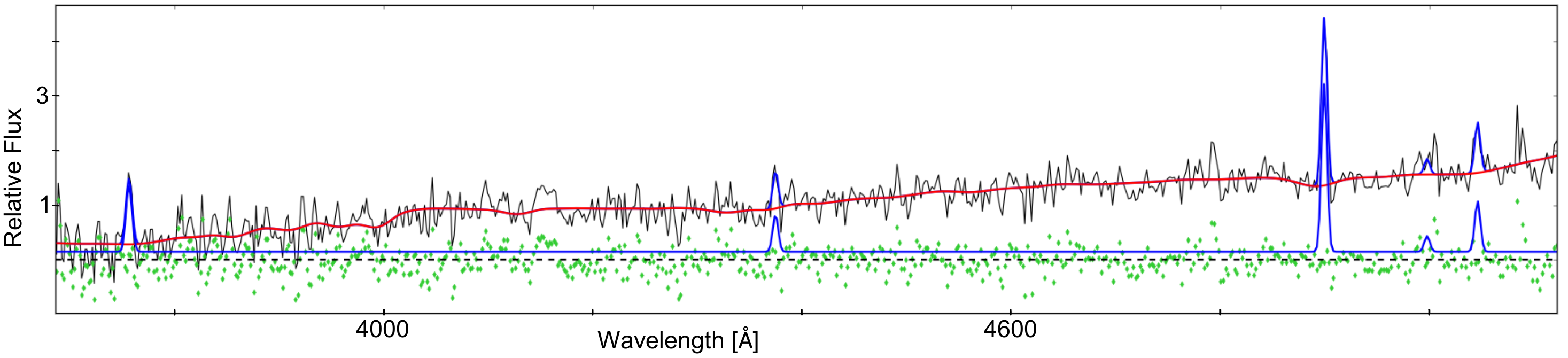}
      \caption{pPXF spectrum of region (2) of NGC~4522 (grism 7): The black line represents the observed spectrum, the blue line shows the emission lines fitted by pPXF to the observed spectrum as well as the extracted pure emission line spectrum.  The red line is the fitted continuum to the data and the green dots are the residuals around the zero value (gray dashed line).}
         \label{ppxf Spektrum}
\end{figure*}

\section{Analysis} 
\subsection{Emission line measurements} 
The spectra were analyzed using the {\tt splot} task in IRAF. We chose the size 
of the extraction region of the extraplanar and disk H{\sc ii} regions depending on 
the extent of their H$\alpha$ emission along the spatial axis in the spectra. 
The emission line flux densities were measured with
the {\tt deblending} tool of IRAF/{\tt splot}, with which we fit gaussians to obtain fluxes
and central peak positions. 
We used the overlapping wavelength range of the two grisms for putting the 
spectra on a common flux scale. Unfortunately, the straight calibration showed
different absolute fluxes of the emission lines due to differences in sky 
transparency conditions.  Thus, it was necessary to first scale the fluxes of
the two grisms to one another. The common lines of H$\beta$ and [OIII]
$\lambda$5007 were used for this in grism 7 and grism 9.

\subsection{Stellar continuum and reddening correction} 
Since absorption lines in the underlying stellar continuum can affect 
the measurement of the emission lines (e.g. the Balmer lines), we need to investigate this effect in our spectra.  
We identified a weak contribution of underlying stellar continuum in the disk H{\sc ii} regions, which was even smaller in the extraplanar 
H{\sc ii} region (1) of NGC~4522 in our spectra. We corrected the H$\alpha$ and 
H$\beta$ lines for absorption with 2 \r{A} in the equivalent width (EW) \citep{moustakas2010}. 
There were no metal-line absorption features visible 
in the stellar continuum of the disk H{\sc ii} regions and therefore it was not possible to 
perform a reliable population synthesis, e.g. with STARLIGHT
\citep{cidfernandesetal2004}. Nevertheless, we tried pPXF fitting
\citep{Capellari2004, Cappellari2017} for one H{\sc ii} region. This routine does full spectral fitting to extract stellar and
gas kinematics, as well as stellar populations information, simultaneously \citep{Cappellari2017}.
It was done for the disk H{\sc ii} region (2) in grism 7 of NGC~4522 to be sure that our applied "standard" absorption 
correction is reliable. As Fig. \ref{ppxf Spektrum} shows, the pPXF-predicted 
absorption features are very weak. The resulting corrected line ratio of 
H$\gamma$/H$\beta$ from pPXF was 0.26, and after   
dereddening 0.29, which is at odds with the theoretical value of 0.46.
Clearly, due to the missing information of the
metal-absorption features, the pPXF code cannot find a good solution for 
our data. Therefore, a first order EW correction of the Balmer lines seems to be 
the best option. 
There was no continuum visible in the extraplanar H{\sc ii} regions (2), (3) of 
NGC~3628 and therefore no equivalent width correction was applied, similar 
to the approach taken for the analysis of faint high redshift galaxies 
\citep[e.g.][]{Mannucci2009}.
To account for extinction, the fluxes were dereddened with the IRAF task {\tt redcorr}. For
this we used the theoretical Balmer decrement of H$\alpha$/H$\beta$~=~2.87
(Case B, 10,000 K), the Galactic extinction law f$(\lambda)$ from \citet{savagemathis1979} and the extinction parameter c (see Table
\ref{table:Hbeta}) with the following equation used in IRAF:
\begin{align}
\text{I}_{corr}(\lambda) = \text{I}_{obs}(\lambda) \cdot 10^{c \ \cdot \text{f}(\lambda)}
\label{eq:balmerdec}
\end{align}
where I$_{corr}(\lambda)$ is the corrected flux and I$_{obs}(\lambda)$ the observed one.
To judge the effect of the chosen 
extinction law, we tested our reddening
correction by using the LMC extinction curve \citep{howarth1983}. For the
H$\alpha$/H$\beta$ line ratio, the change is 0.3\%, whereas the biggest change
is 3.5\% with the [OII]/H$\beta$ line ratio. We claim the reddening
correction is only weakly dependent on the adopted reddening curve. If possible, the lines
were measured in grism 9. \\ \indent
The H$\beta$ fluxes are listed in Table
\ref{table:Hbeta}. The dereddened and scaled fluxes relative to H$\beta$ are
listed in Table \ref{table:2}.

\begin{table*}
\caption{H{\sc ii} region properties.}             
\label{table:Hbeta}      
\centering          
\begin{tabular}{c l c c c c c}    
\hline\hline       
Galaxy & Region   & H$\beta$ grism 7                      & H$\beta$ grism 9  &                      c     & v$_{helio}$ [$\pm$ 23 km/s]   &  size H{\sc ii} region [pc]  \\ 
\hline                 
   NGC~3628 & 1    & (4.63 $\pm$ 0.67) $\times$ 10$^{-16}$ & (4.31 $\pm$ 0.40) $\times$ 10$^{-16}$ & 0.95  &   671    &    \\ 
            & 2 (e)& (3.89 $\pm$ 0.56) $\times$ 10$^{-16}$ & (3.89 $\pm$ 0.36) $\times$ 10$^{-16}$ & \ \ \ \ 0.12\tablefootmark{(1)}  &   648    &  150 $\pm$ 30  \\ 
            & 3 (e)& (2.53 $\pm$ 0.36) $\times$ 10$^{-16}$ & (2.53 $\pm$ 0.23) $\times$ 10$^{-16}$ & \ \ \ \ 0.00\tablefootmark{(2)} &   650    &  150 $\pm$ 30 \\ 
   NGC~4522 & 1 (e)& (6.22 $\pm$ 0.90) $\times$ 10$^{-16}$ & (6.12 $\pm$ 0.56) $\times$ 10$^{-16}$ & 0.30  &   2366   &  325 $\pm$ 40 \\
            & 2    & (1.22 $\pm$ 0.18) $\times$ 10$^{-15}$ & (1.35 $\pm$ 0.12) $\times$ 10$^{-15}$ & 0.33  &   2320   &    \\
            & 3    & (5.67 $\pm$ 0.82) $\times$ 10$^{-16}$ & (6.04 $\pm$ 0.56) $\times$ 10$^{-16}$ & 0.37  &   2253   &    \\
\hline                  
\end{tabular}
\tablefoot{H$\beta$ fluxes in erg $cm^{-2}$ s$^{-1}$; (e) $=$ extraplanar.\\
\tablefoottext{1}{no EW correction applied}, \tablefoottext{2}{no EW correction and no reddening correction applied}
}
\end{table*}

\begin{table*}
\caption{Dereddened and scaled fluxes normalized to H$\beta$.}             
\label{table:2}      
\centering          
\begin{tabular}{c l c c c c c c c c c c c c c c c c}    
\hline\hline       
Galaxy & Region   &  [OII] $\lambda\lambda$ 3726, 3728 &  H$\gamma$ & [OIII] $\lambda$4959 & [OIII] $\lambda$5007 & [NII] $\lambda$6548 &H$\alpha$      & [NII] $\lambda$6583\\ 
\hline                 
   NGC~3628 & 1    & 1.85 $\pm$ 0.20       					& 0.48 $\pm$ 0.09 &  0.14 $\pm$ 0.04   & 0.31 $\pm$ 0.05      & 0.27 $\pm$ 0.04    & 2.87 $\pm$ 0.31 & 0.82 $\pm$ 0.12\\ 
            & 2 (e)& 1.11 $\pm$ 0.16                & 0.46 $\pm$ 0.09 &  0.52 $\pm$ 0.11   & 1.57 $\pm$ 0.25      & 0.16 $\pm$ 0.03    & 2.87 $\pm$ 0.31 & 0.50 $\pm$ 0.07\\ 
            & 3 (e)& 1.71 $\pm$ 0.32  							& 0.40 $\pm$ 0.11 &  -							   & 0.30 $\pm$ 0.05      & 0.17 $\pm$ 0.03    & 2.84 $\pm$ 0.31\tablefootmark{(1)} & 0.54 $\pm$ 0.08\\ 
   NGC~4522 & 1 (e)& 2.02 $\pm$ 0.32 								& 0.45 $\pm$ 0.11 &  -  							 & 0.36 $\pm$ 0.08      & 0.17 $\pm$ 0.04    & 2.87 $\pm$ 0.31 & 0.70 $\pm$ 0.10\\
            & 2    & 0.88 $\pm$ 0.14  							& 0.40 $\pm$ 0.10 &  0.24 $\pm$ 0.03   & 0.38 $\pm$ 0.07      & 0.28 $\pm$ 0.04    & 2.87 $\pm$ 0.31 & 0.95 $\pm$ 0.13\\
            & 3    & 0.89 $\pm$ 0.16 							  & -               &  0.09 $\pm$ 0.01   & 0.31 $\pm$ 0.05      & 0.35 $\pm$ 0.06    & 2.87 $\pm$ 0.31 & 0.85 $\pm$ 0.13\\
\hline                  
\end{tabular}
\tablefoot{(e) $=$ extraplanar, \tablefoottext{1}{no dereddening applied}.}
\end{table*}
 
\subsection{Errors}
\subsubsection{Flux errors}
Errors in the emission line fluxes can arise from the calibration and from the measurement of the emission lines with the task {\tt splot} in IRAF. Statistical errors are calculated within {\tt splot}.
\begin{itemize}
\item The error in the flux calibration was tested by applying the calibration to a standard star and comparing the resultant fluxes with the known ones. The error was found to be 12\% in grism 7 and 6\% in grism 9.
\item The task {\tt splot} itself displayed errors after specifying the parameters inverse gain and noise $\sigma_0$. For each emission line, the error was determined separately. We tested these errors with the [OII] $\lambda\lambda$ 3726, 3728 and [OIII] $\lambda$5007 lines in grism 7 and with H$\alpha$ and [OIII] $\lambda$5007 lines in grism 9. By repeated measurements of the same emission line with variations in applying the {\tt deblending} tool within the {\tt splot} task, for e.g. variation in the initial estimate of the peak position and continuum, the resulting errors were found to be similar to the {\tt splot} ones. Therefore the {\tt splot} errors were used for further error propagation.
\end{itemize}
The resulting flux error (sum of the squares) for H$\beta$ is 14.4\% in grism 7 and 9.2\% in grism 9. The errors are propagated throughout the complete abundance analysis (gaussian error propagation). To calculate the final error in the oxygen abundance determination, we added the errors of the different calibrations to the flux error (sum of squares).
\subsubsection{Wavelength errors}
The errors of the velocity determination result from the wavelength calibration and the measurement of the peak positions of the emission lines.
\begin{itemize}
\item The applied wavelength calibration function, had a dispersion of 0.5~\r{A}, which was taken as the error of the calibration. This was further checked by measuring the position of three sky lines per grism and comparing them with values in the literature \citep{himmelslinien2003}. The error was confirmed to be 0.5~\r{A}.  
\item The error of the peak positions was determined by measuring four times the same emission line with variation in the initial estimate of the peak position using the {\tt deblending} tool within the {\tt splot} task. The average of the errors given by {\tt splot} was taken as the resulting error.
\end{itemize}
The total error in wavelength is 2.13~\r{A} in grism~7, which mainly originates from the high {\tt splot} errors due to the low S/N in this grism. The total wavelength error is 0.50~\r{A} in grism~9.  
\subsection{Parameter determination}  
The projected distances from the extraplanar regions to the host galaxy were determined as the projected height perpendicular to the central line of the disk, which was defined by an isophotal fit to the 3.6~$\mu$m image for NGC 4522 and by the central emission in the H$\alpha$ image for NGC 3628.
The heliocentric velocities of the H{\sc ii} regions were determined from the redshifts of the H$\alpha$ line, which is the brightest line. The heliocentric correction was calculated by the IRAF task {\tt rvcorrect}.\\ \indent
The metallicities were derived using the oxygen abundance 12~+~log(O/H) by the following three different methods. 
\subsubsection{Method 1}
\cite{kewleydopita2002} (theoretical) use stellar population synthesis and photoionization models to produce a set of ionization parameter and abundance diagnostics based on the use of the strong optical emission lines. They claim the ratio of [NII] $\lambda$6583 to [OII] $\lambda\lambda$3726, 3729 is a reliable diagnostic because it is independent of the ionization parameter and has no local maximum like the R$_{23}$ parameter. Additionally, the [NII], [OII] lines are quite strong, even in low S/N spectra. With R~=~I($\text{[NII]}\lambda$6583)/I($\text{[OII]}\lambda\lambda$3726, 3729) a first estimate of the oxygen abundance is calculated:
\begin{align}
& 12~+~\text{log(O/H)} = \notag \\ 
 & \ \ \ \ \ \ \ \ \ \ \ \ \ \text{log}(1.54020 + 1.26602 \text{R} + 0.167977 \text{R}^2) + 8.93 
 \label{eq:m1}
\end{align}
Following that paper, the oxygen abundance is taken as final if the result is $12~+~\text{log(O/H)} \geq$ 8.6, which was the case in all regions discussed in this paper.\\
\indent
\textit{Limitations:}
\cite{kewleydopita2002} mention that the observed [NII]/[OII] abundance ratio shows large scatter from object to object due to the varying age distribution of the stellar population for a sample of H{\sc ii} regions. Furthermore, in this method, the oxygen abundance depends only on the [NII]/[OII] ratio, which is taken from an empirical fit to the observed behavior of the [NII]/[OII] ratio in H{\sc ii} regions. Because the lines are well separated in wavelength, this method uses one line from each grism and hence it is highly dependent on the scaling and the reddening correction. \\
\indent We also used the ionization parameter, which was determined empirically \citep{kewleydopita2002}:
 \begin{align}
 \text{log(I([OIII])}& /\text{(I([OII]))} =  \notag \\
& \text{k}_0 +  \text{k}_1  \text{log(q)} + \text{k}_2  \text{log(q)}^2 + \text{k}_3  \text{log(q)}^3  
 \label{eq:m1b}
\end{align}
with given coefficients for 12~+~log(O/H) = 8.9 $\text{k}_0$ = -52.6367, $\text{k}_1$ = 16.088, $\text{k}_2$ = -1.67443 and $\text{k}_3$ = 0.0608004 

\subsubsection{Method 2}
\citet{pilyuginetal2014} is a revised version of \cite{pilyugin2001,pilyuginthuan2005}. They calibrated empirically the R$_{3}$ and P (excitation parameter) against the oxygen abundances determined with temperature-sensitive lines (direct T$_e$ method) of more than 250 H{\sc ii} regions from 3904 different spectra of 130 late-type galaxies, providing central oxygen abundances and abundance gradients. This is done for low metallicities (12~+~log(O/H) $\leq$ 8.0) \citep{pilyugin2000} and high metallicities (12~+~log(O/H) $\geq$ 8.3). In this paper the relation for the higher metallicities (upper branch: $12~+$~log(O/H) $\geq$ 8.3) is used. \\
\indent
The excitation parameter is defined as P~=~R$_3$/R$_{23}$ whereas 
\begin{align}
							&		\text{R}_3~=~\text{I([OIII]} \lambda\lambda4959, 5007)/\text{I(H}\beta)\\
\label{eq:m2} &  \text{R}_{23} = \frac{\text{I([OII]} \lambda\lambda3726, 3729) + \text{I([OIII]} \lambda\lambda4959, 5007)}{\text{I(H}\beta)}\\ 
\end{align}

Using this definition of the excitation parameter P and R$_3$, they computed:
\begin{align}
\label{eq:m2c}  12~+~\text{log(O/H)} &= 8.334 + 0.533 \text{P} \notag \\
																			&- (0.338 + 0.415 \text{P}) \text{log R}_3 \notag \\
																			&- (0.086 - 0.225 \text{P})(\text{log R}_3)^2 
\end{align}
With equation (\ref{eq:m2c}) the oxygen abundance was determined.\\
\indent
\textit{Limitations:}
In the intermediate-metallicity regime, $12~+$~log(O/H)~$\sim$~8.0 - 8.4, the derived oxygen abundance is highly affected by the R$_{23}$ ambiguity, therefore the average value of the higher and lower metallicity regime should be assumed \citep{lopezetal2012}.

\subsubsection{Method 3} 
\cite{pettinipagel2004} empirically calibrated the parameter O3N2 (eq. \ref{eq:m3}) against the oxygen abundances of a sample of 137 extragalactic H{\sc ii} regions determined mostly from temperature-sensitive lines (direct T$_e$ method) or photoionization models. This results in the oxygen abundance determination via a linear fit between -1 and 1.9, which is used in this paper (eq. \ref{eq:m3b}). 
\begin{align}
 \label{eq:m3} & \text{O3N2} = \text{log}\left(\frac{\text{I}(\text{[OIII] }\lambda 5007)/\text{I}(\text{H}\beta)}{\text{I}(\text{[NII] }\lambda6583)/\text{I}(\text{H}\alpha)}\right) \\ 
 \label{eq:m3b} & 12~+~\text{log(O/H)} = 8.73 - 0.32 \cdot \text{O3N2}
\end{align}
\indent
\textit{Limitations:}
This method is only valid for $12~+~\text{log(O/H)}~\gtrsim$~8.7 \citep{lopezetal2012}.

\section{Results and discussion}
Tables \ref{pn} and \ref{parameter} show the results of the analysis. In this section, we discuss the Balmer decrement, compare the calibrations and describe the results of the individual galaxies and their H{\sc ii} regions.
\begin{table}
\caption{[OIII]/H$\beta$ and ionization parameter q of NGC~3628.}             
\label{pn}      
\centering     
	\begin{tabular}{c l c c}
	\hline\hline		
Galaxy& Region & [OIII]/ H$\beta$  & q \citep{kewleydopita2002}               \\
\hline 
NGC~3628 &1    & 0.31 $\pm$ 0.05  & (1.95 $\pm$ 0.23) $\times$ 10$^7$              \\
         &2 (e)& 1.57 $\pm$ 0.25  & (2.68 $\pm$ 0.30) $\times$ 10$^8$             \\
         &3 (e)& 0.30 $\pm$ 0.05  & (1.67 $\pm$ 0.16) $\times$ 10$^7$           \\
\hline
\end{tabular}
\tablefoot{(e) $=$ extraplanar}
\end{table}

\begin{table*}
\caption{Resulting oxygen abundances of the different methods.}             
\label{parameter}      
\centering     
	\begin{tabular}{c l c c c}
	\hline\hline		
       &         &    \multicolumn{3}{c}{12~$+$~log(O/H)}                                 \\
       \cline{3-5}
Galaxy& Region & Method 1              &  Method 2            &        Method 3                    				    \\
       &     & \citet{kewleydopita2002} & \citet{pilyuginetal2014}& \citet{pettinipagel2004}       \\
       &				&			[$\pm$ 0.04]		&		[$\pm$ 0.10]				&	[$\pm$ 0.14]		\\
\hline 
NGC~3628 &1    & 8.96 $\pm$ 0.06   & 	8.58 $\pm$ 0.11	  		&	8.72 $\pm$ 0.15             					         \\
         &2 (e)& 8.94 $\pm$ 0.07   & 	8.43 $\pm$ 0.14	  		&	8.42 $\pm$ 0.21              					       \\
         &3 (e)& 8.92 $\pm$ 0.07   & 	8.60 $\pm$ 0.11				&	8.67 $\pm$ 0.15              					      \\
 \hline
NGC~4522 &1 (e) & 8.92 $\pm$ 0.07  & 	8.54 $\pm$ 0.13				&		8.63 $\pm$ 0.16					  					  	\\
         &2     & 9.13 $\pm$ 0.04  & 	8.68 $\pm$ 0.12				&		8.69 $\pm$ 0.15											   \\
         &3     & 9.12 $\pm$ 0.04  & 	8.69 $\pm$ 0.11	  		&		8.72 $\pm$ 0.15					 						   \\
 \hline
\end{tabular}
\tablefoot{(e) $=$ extraplanar, the error in the heading is the intrinsic error of the calibration according to the publication.}
\end{table*}
\begin{figure*}
\begin{center}
\sidecaption
{\resizebox{11.5cm}{!}{\includegraphics{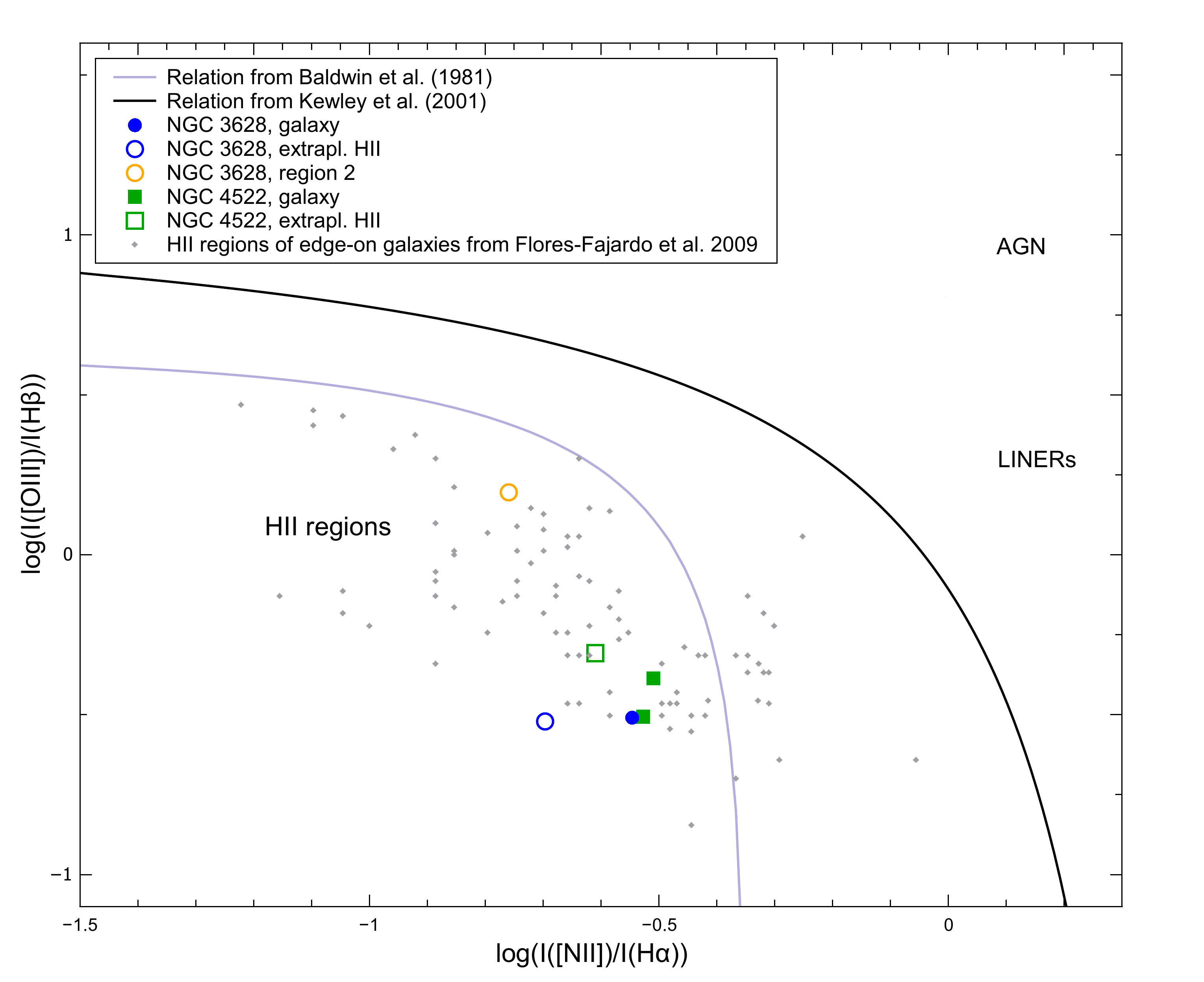}}}
\caption{Diagnostic diagram: The lines are a fit through H{\sc ii} regions from \citet[][grey]{baldwinetal81} and the theoretical upper limit for the starburst
model from \citet[][black]{kewley2001}. The errors of the six H{\sc ii} regions correspond to the size of the symbols. The grey diamonds are H{\sc ii} regions from a sample of egde-on galaxies from \citet{floresfajardo2009}.}
\label{Diagnostic Diagramm}
\end{center}
\end{figure*}

\subsection{Balmer decrement of the H{\sc ii} regions}
\label{4.2}
The measured Balmer decrement of H$\alpha$/H$\beta$ from the extraplanar H{\sc ii} region (3) of NGC~3628 is slightly lower than the theoretical value. Possibly, the region has a higher temperature. Nevertheless, considering the error, the Balmer decrement fits the theoretical value for all regions. The extinction parameter of the extraplanar H{\sc ii} region (3) of NGC~3628 is zero. Therefore, no reddening correction was applied. As the two extraplanar regions (2), (3) of NGC~3628 are away from the galaxy disk they are expected to be only little influenced by reddening, which is consistent with our observation. \\ \indent
We did a sanity check on region (3) of NGC~3628 to test how sensitive the oxygen abundance is to the  uncertainty in [OII] $\lambda\lambda$ 3726, 3728. The [OII] line in grism 7 is very important for method 1, however grism 7 does not provide high S/N. We determined the influence of this line on the abundance measurement. We calculated the oxygen abundance with method 1 using the measured value of H$\alpha$/H$\beta$ = 2.84 and additionally with considering the error of 0.31 which leads to the upper limit of the H$\alpha$/H$\beta$ ratio of 3.15. The latter ratio would lead to an extinction parameter of c = 0.13. That increases the [OII] $\lambda\lambda$ 3726, 3728 line ratio from 1.71 to 1.86 and decreases [NII] $\lambda$6583 from 0.54 to 0.49. The resulting oxygen abundance of method 1 would change from 8.92 to 8.87. Thus, the oxygen abundance determination is robust towards our errors and the analysis is therefore reliable.

\subsection{Oxygen abundances - general trends and comparison of the calibrations} 
\label{4.3}
The results of the abundance determination are consistent within all regions.\\ \indent
In general, method 1 \citep{kewleydopita2002} shows the highest values of 12~+~log(O/H) while method 2 \citep{pilyuginetal2014} and method 3 \citep{pettinipagel2004} provide lower values.  The mean discrepancy between method 1 and method 2 is 0.41 dex. The mean offset between method 1 and method 3 0.36 dex and between method 2 and method 3 0.05 dex. This result is in agreement with other analyses \citep{modjazkewley2008, kewleyellison2008}.\\
\indent To check the determined oxygen abundances, we searched for reference metallicities of the galaxies. 
For NGC~3628, an oxygen abundance of 12~+~log(O/H)~=~8.57 was determined by  \citet{engelbrachtetal2008} using an integrated spectrum of the galaxy and empirical strong-line methods \citep{pettinipagel2004, pilyuginthuan2005}. To compare to their value, we calculated also the mean of the methods. For the disk H{\sc ii} region~(1) of NGC~3628 the reference oxygen abundance is similar to the mean values of the two calibrations of 8.56 dex.  \\
\indent The comparison of theoretical and empirical strong-line calibrations show well known offsets \citep{stasinska2002, modjazkewley2008, moustakas2010}. We confirm the offsets with a mean difference of 0.37 dex. \\ 
\indent Strong-line calibrations use the fact that the metallicity in large H{\sc ii} regions is connected to the mean effective temperature of the stellar radiation field and the ionization parameter \citep{stasinska2010}. The theoretical calibrations use photoionization models to derive metallicities. These models are limited and depend strongly on the temperature structure calculations of the nebula \citep{kewleyellison2008}. This leads to discrepancies of up to 0.2 dex within the theoretical calibrations and up to 0.6 dex in comparison to empirical methods. Empirical calibrations use the relationship between direct metallicities and strong-line ratios of a sample of H{\sc ii} regions. This sample shows special characteristics in densities, the strength of the radiation field, and the state of ionization \citep{stasinska2010}. Therefore, empirical methods should be used with H{\sc ii} regions which are comparable to the calibration sample. One origin of discrepancies within different empirical methods is the variety of different characteristics of H{\sc ii} regions. A possible reason for the offset in the comparison to theoretical calibrations are temperature fluctuations in H{\sc ii} regions. In metal-rich H{\sc ii} regions effective cooling processes could lead to strong temperature gradients \citep{garnett92}. Therefore the temperature determined with the forbidden lines is overestimated and the oxygen abundance underestimated \citep{garnett92}. With knowledge of the fluctuations, the oxygen abundance would shift 0.2 -- 0.3 dex \citep{moustakas2010}. Another possible reason has been mentioned by \citet{nichollsetal2012}. If the electrons in H{\sc ii} regions are not following a Maxwell-Boltzmann equilibrium energy distribution but a "non-equilibrium kappa distribution", it should be possible to estimate the temperature and metallicity in these regions more accurately.\\
\indent In addition to the calibrations used in this paper,  \citet{kewleyellison2008} also investigated seven other calibrations and got a mean difference of up to 0.7 dex between theoretical and empirical calibrations. They propose to use oxygen abundance calibration with caution and to always specify the adopted calibration. They refer to relative oxygen abundances as a reliable tool, especially with the calibrations of \citet{kewleydopita2002} and \citet{pettinipagel2004}. 

\subsection{Individual galaxies}
In this section the individual galaxies with their extraplanar H{\sc ii} regions are presented. In
interpreting our results, we will take account of the metallicity gradients in spiral galaxies \citep[e.g.,~][]{garnett1998,sanchezetal2016} which can often change the metallicity by up to 0.3 dex within R$_{25}$ and does not depend on the different metallicity calculations \citep[e.g.,~][]{magrinietal2011}.
 
\subsubsection{NGC~3628}
NGC~3628 is a member of the Leo triplet and interacts with NGC~3627. The two analyzed extraplanar H{\sc ii} regions are located in a filament of NGC~3628 visible in HI \citep{wilding93}, which was probably formed due to tidal interactions. The regions have projected distances of 2.8 kpc (2) and 3 kpc (3).\\ \indent
\textit{Radial velocity:} The determined radial velocities of the extraplanar H{\sc ii} regions are 648 km/s of region (2) and 650 km/s of region (3). Comparing this result to an HI velocity map of \cite{wilding93}, the velocities are consistent. The same is true for the disk H{\sc ii} region (1). It is also obvious that the analyzed extraplanar regions (2), (3) and the disk region (1) are located in the same section of the velocity field. The extraplanar regions of NGC~3628 are located in a filament, which was probably formed due to interaction with NGC 3627. The velocity gradient within the filament as seen in the HI map is a hint that it has its origin in the outer parts of the galaxy. Therefore, the extraplanar regions and the disk region have comparable velocities within the rotation curve. The measured heliocentric velocities from our spectra confirm this (Table~\ref{table:Hbeta}). There is a difference of only 20 km/s between the disk region and the extraplanar regions. Therefore, the extraplanar regions are comparable with the disk region in terms of metallicity gradients. \\ \indent
\textit{Metallicities:}
The derived oxygen abundances of the H{\sc ii} disk region (1) and extraplanar region (3) are the same within the errors in all methods. In method 1 the extraplanar region (3) has a slightly lower value than the disk region (1). Nevertheless, in general, region (3) seems to have a comparable abundance to the disk region (1). We conclude that this extraplanar region was formed in a tidal arm of disk material. \\ 
\indent Region (2) of NGC~3628 shows the lowest value in method 3. This value is confirmed by method 2. There is an indication that this region is highly ionized, based on the high [OIII]/H$\beta$ ratio and the high ionization parameter in comparison to the other H{\sc ii} regions of this galaxy (see Table \ref{pn}). In the diagnostic diagram (Fig. \ref{Diagnostic Diagramm}) the region shows the highest value of log(OIII/H$\beta$) of the H{\sc ii} regions analyzed in this paper but is still within the region of H{\sc ii} regions located. 

\subsubsection{NGC~4522}
NGC~4522 is a member of the Virgo cluster. The optical distribution is undisturbed \citep{kenneykoopmann99} while the HI distribution is asymmetric \citep{kenney2004}. There is indication that ram pressure of the Virgo cluster stripped the gas of the galaxy \citep{kenney2004}. The extraplanar H{\sc ii} region (1) is located in the outer gas and has a projected distance of 1.4 kpc.\\ \indent
\textit{Radial velocity:} The determined heliocentric velocities of the three H{\sc ii} regions are around 2310 km/s. The extraplanar region (1) differs by 46 km/s from disk region (2) and by 113 km/s from disk region (3) in velocity (Table~\ref{table:Hbeta}). A similar trend is seen in the velocity map of \citet{kenney2004} at the locations of the regions, but with values that are, on average, lower by 30 km/s.  One reason is that their central heliocentric velocity of 2337 km/s is offset to other measurements, which are lower e.g. from HI of \citet{vaucouleurs1991} at 2324~$\pm$~5 km/s. As the H$\alpha$ line lies at the edge of grism 9, we tested our measured heliocentric velocity using the [OIII] $\lambda$5007 line, which  lies in the middle of grism 9. We find on average 15 km/s higher heliocentric velocities, more in
agreement  with \citet{kenney2004} and indeed 
 compatible within the errors.\\ \indent
\textit{Metallicities:} The extraplanar H{\sc ii} region (1) shows lower oxygen abundances in comparison to the disk H{\sc ii} regions (2) and (3) in all methods, with the mean difference being 0.17 dex. In method 3 the offset between disk and extraplanar H{\sc ii} regions is the smallest and not significant, whereas the offset of 0.20 dex in method 1 is the largest and most significant. Method 2 shows a slightly lower value. With considering the combined errors this is not significant either. However, considering all methods together the lower metallicity is slightly more significant than for method 1 alone. In the diagnostic diagram (Fig. \ref{Diagnostic Diagramm}), all regions are located close to each other. The Virgo Cluster galaxy NGC~4522 is close to peak ram pressure, where the outer gas disk has already been removed and the remaining gas disk is strongly truncated \citep{vollmer2009}. The slightly smaller oxygen abundance of the extraplanar H{\sc ii} region could be caused due to the metallicity gradients which decrease the metallicity in the outer parts, which are the regions most affected by ram pressure stripping. We conclude that the extraplanar H{\sc ii} region (1) is located in these parts of stripped gas from the galaxy and therefore shows slightly lower metallicities.
\subsubsection{Origin of extraplanar H{\sc ii} regions}
The three analyzed extraplanar H{\sc ii} regions show different origins. These are tidal interactions in the galaxy NGC~3628 (tidal stream) and ram pressure in NGC~4522. They both are influenced by the group/cluster environment. As the extraplanar H{\sc ii} regions studied here show comparable or slightly lower oxygen abundances in comparison to the H{\sc ii} regions in the host galaxy, they likely result from disk material. This result is different from the analysis on NGC~55 \citep{tuellmann2003, kudritzkietal2016} where two analyzed extraplanar H{\sc ii} regions show low oxygen abundance. Therefore, ejection from the disk was ruled out and they claimed that the regions were formed in the halo. The intergalactic regions of the Stephan's Quintet \citep{mendesetal2004} show a high oxygen abundance. These regions were probably formed in tidal tales of the four interacting galaxies. The extragalactic region of NGC~4402 \citep{corteseetal2004} seems also to have a high metallicity and thus is built with pre-enriched material. They ruled out a tidal stripping scenario due to the absence of a clear interaction signal. One possible scenario would be that the extragalactic H{\sc ii} region is formed by an interaction between NGC~4402 and the ICM of the Virgo. The mechanism of ram pressure stripping is probably present in the region of NGC~4388 \citep{oosterloovangorkom2005}. The same is true for the extraplanar H{\sc ii} region (1) of NGC~4522 analyzed in this paper, whereas the metallicity is low in the extraplanar region of NGC 4388 \citep{gerhardetal2002} and just slightly lower in our case.

\section{Summary and Conclusions}

We present optical long-slit spectroscopic data of extraplanar H{\sc ii} regions
and their edge-on host galaxies NGC~3628 and NGC~4522 in this paper. In order
to analyze these extraplanar H{\sc ii} regions, we determined metallicities (oxygen
abundances). As the spectra contain strong emission lines like the Balmer
series, [OII] $\lambda\lambda$3726, 3729, [OIII] $\lambda\lambda$4959, 5007,
and [NII] $\lambda\lambda$6548, 6583, we calculate the oxygen abundances
of the extraplanar H{\sc ii} regions with different strong-line calibrations (based on empirical data and on theoretical photoionization grids).
First, we probed the shift between the different types of
strong-line calibrations (empirical and theoretical). We analyzed the
extraplanar H{\sc ii} regions as well as at least one H{\sc ii} region located in
the disk of the galaxies by using three different calibrations
\citep{kewleydopita2002, pilyugin2001, pilyuginetal2014, pettinipagel2004}. Our analysis again
confirms the known offsets with a mean value of 0.37 dex.
Secondly, we investigated the origin of the extraplanar H{\sc ii} regions by
comparing the oxygen abundance of the extraplanar and disk H{\sc ii} regions.\\ \indent
In NGC 3628 we derived comparable oxygen abundances, which leads to the conclusion that the
  extraplanar H{\sc ii} regions are formed from disk material. The two analyzed
extraplanar H{\sc ii} regions (2), (3) of NGC~3628 are located in a filament that emerges from the
disk, which was probably formed due to tidal interactions with NGC~3627.\\ \indent
The extraplanar H{\sc ii} region (1) of NGC~4522 shows slightly lower measured
oxygen abundances in comparison to the disk H{\sc ii} regions (2), (3). Therefore, ram pressure of the Virgo cluster probably stripped out the gas of the
  galaxy. This outer gas, where the extraplanar H{\sc ii} region is located, is not
  as metal rich as the inner disk region as a result of metallicity gradients. Thus, this extraplanar H{\sc ii} region (1) shows a slightly lower oxygen abundance than the inner disk.\\ \indent 
Taking our results together with our previous work on NGC~55 \citep{tuellmann2003} and the the results of other groups (see section 1),
it is clear that the mechanisms for forming extraplanar regions are diverse. The H{\sc ii} regions
analyzed in this paper, as well as most of the extraplanar H{\sc ii} regions
mentioned in the literature, do not result from star formation in low-metallicity halo
material. The extraplanar H{\sc ii} regions seem largely to result from material pulled out during an 
interaction or due to ram pressure stripping. Therefore, the environment
has a great impact on the generation of H{\sc ii} regions far above the disk plane. \\ \indent 
We conclude that infalling gas, which is probably composed of falling back disk material and a small
amount of intergalactic medium \citep[e.g.][]{sancisietal2008}, is not
forming stars in the halo very frequently.

\begin{acknowledgements}
We thank the anonymous referee for constructive comments to improving this work. This research was suported in part by the DFG (German Research Foundation) research unit FOR1048, has made use of IRAF, the NASA's Astrophysics Data System Bibliographic Services, the GOLDMine Database and the NASA/IPAC Extragalactic Database (NED) which is operated by the Jet Propulsion Laboratory, California Institute of Technology, under contract with the National Aeronautics and Space Administration.
\end{acknowledgements}

\bibliography{Bibliography}
\bibliographystyle{aa}

\end{document}